\documentclass[a4paper,11pt]{article}

\newcommand\ion[2]{\text{#1\,\textsc{\lowercase{#2}}}}	

\def\emerlin{$e$-MERLIN}

\usepackage{pos}
\usepackage{wrapfig}

\title{The LeMMINGs survey: probing sub-kpc radio structures of nearby galaxies with e-MERLIN}

\manuallySeparateAuthors
\author*[a]{David R. A. Williams}
\author[b]{, Ranieri D. Baldi}
\author[a]{, Robert J. Beswick (co-PI)}
\author[c]{, Ian M. McHardy (co-PI)}
\author[d]{, Bililign T. Dullo}
\author[e]{, Mayukh Pahari}
\author[c]{, Emma Carver}
\author[c]{, Jake Clifford}
\author[c]{, Nicholas A. Kill}
\author[c]{, Bhairavi Krishnamoorthi}
\author[a]{, Oliver Woodcock}
\author[f,g]{, Johan Knapen}
\author[h,i]{, Smita Mathur}
\author{on behalf of the LeMMINGs collaboration}

\affiliation[a]{Jodrell Bank Centre for Astrophysics, The University of Manchester, \\
  School of Physics and Astronomy, The University of Manchester, Manchester, M13 9PL, UK}

\affiliation[b]{INAF - Istituto di Radioastronomia,\\ I-40129 Bologna, Italy }

\affiliation[c]{School of Physics and Astronomy, University of Southampton,\\
Southampton, SO17 1BJ, UK}

\affiliation[d]{Departamento de F\'isica de la Tierra y Astrof\'isica, IPARCOS, Universidad Complutense de Madrid,\\
E-28040 Madrid, Spain}

\affiliation[e]{Department of Physics, Indian Institute of Technology,\\
Hyderabad 502285, India}

\affiliation[f]{Instituto de Astrof\'{i}sica de Canarias, V\'{i}a L\'{a}ctea S/N,\\
La Laguna, E-38205, Spain}

\affiliation[g]{Departamento de Astrof\'{i}sica, Universidad de La Laguna,\\
La Laguna, E-38205, Spain}

\affiliation[h]{Astronomy department, The Ohio State University,\\
Columbus, OH, 43210, USA}

\affiliation[i]{Center for Astronomy and Astro-particle Physics, The Ohio State University,\\
Columbus, OH 43210, USA}

\emailAdd{david.williams-7@manchester.ac.uk}
\emailAdd{}

\abstract{The Legacy \emerlin{} Multi-band Imaging of Nearby Galaxies survey (LeMMINGs) is a statistically-complete census of nuclear accretion and star formation processes in the local Universe. The LeMMINGs observations at 1.5 and 5 GHz yield angular resolutions on 10s milliarcsecond-scales, with sensitivities of 10s $\mu$Jy. Awarded 810 hours of observing time, the full statistical sample (at 1.5 GHz) plus several studies of individual objects have now been published. Combined with multi-wavelength follow up observations, this survey will provide a unique legacy data set of our Galactic back yard. We present an overview of the LeMMINGs results so far, including the 1.5\,GHz sample results and associated \textit{Chandra} X-ray data. We describe the next steps for LeMMINGs to analyse the 5\,GHz survey and produce widefield images to categories all radio sources in the LeMMINGs galaxies.}

\FullConference{%
  15th European VLBI Network Mini-Symposium and Users' Meeting (EVN2022)\\
  11-15 July 2022\\
  University College Cork, Ireland
}

\begin{document}
\maketitle

\section{Introduction} 
\vspace{-0.3cm}

Our cosmic back yard hosts a wide variety of galaxies of different sizes and morphologies. These galaxies have been studied across the electromagnetic spectrum over many decades, detecting the presence of accreting super-massive black holes (SMBHs), known as active galactic nuclei (AGN), as well as observing the stellar life cycles from star formation (SF) to supernova remnants (SNRs). Nearby galaxies (d$<$150\,Mpc) allow us to study the end points of galaxy evolution while probing the SF and AGN structures at an unprecedented resolution. Only by using large multi-wavelength statistical samples of nearby galaxies is it possible to understand the connection between these processes and how they shape galaxies over time.

The radio waveband offers a largely unobscured insight into the activity and processes of some of the most energetic phenomena in the Universe. Centimetric wavelength observations are capable of detecting the synchrotron emitting jets of low-luminosity active galactic nuclei (LLAGN) as well as the expanding shells of SNRs. However, the relative contribution of radio emission to the total galaxy emission is quite low, necessitating high sensitivity (i.e. 10s $\mu$Jy) observations to detect these structures. Such structures are apparent on sub-kpc-scales, but can be confused without the requisite sub-arcsecond resolution. 
In nearby galaxies, we can probe linear resolutions of 100 parsecs per arcsecond at a distance of 20\,Mpc. The variety of nearby galaxies that we observe (e.g. elliptical, spiral, irregular, lenticular and dwarf) provides excellent laboratories for SF processes and accretion, enabling population studies of a large sample of all types of galaxy. 

While high resolution, high sensitivity radio observations of nearby galaxies can detect the compact radio sources associated with SF and AGN activity, categorising them is challenging. Use of multiple radio frequencies enables a radio spectral index to be obtained and determine whether the radio emission is thermal or non-thermal in nature. Multi-wavelength (i.e. X-ray/optical/infra-red) studies of matching resolution are also crucial in determining the SF rate, prevalence of high energy X-ray sources and galaxy-wide ionisation processes. 
In order to get the most of any nearby galaxy sample, it must be statistically-complete, i.e. it represents the population to be tested, have good coverage at commensurate high resolutions across the electromagnetic spectrum and be large enough to be statistically meaningful. This is why the LeMMINGs programme exists, with two main science goals: i) quantifying SF rates and the supernova remnant population and ii) to produce a census of accretion activity including radio luminosity functions, in nearby galaxies.

\vspace{-0.3cm}
\subsection{Star formation rates and source populations in nearby galaxies}
While the LeMMINGs sample encompasses all types of galaxy morphologies, it is dominated by star-forming spirals. SF and supernovae (SNe) rates for most of the nearest LeMMINGs galaxies will be measured from the detection of radio SNe and SNRs. Targeted deep observations will enable the detection of fainter radio sources associated with SF such as 
{\text{H\,\textsc{\lowercase{II}}}} regions, planetary nebulae and star clusters. The compact radio source population in these galaxies will enable the calibration of the SF rate in nearby galaxies unobscured by dust, which affects many other SF rate indicators. 

\vspace{-0.3cm}
\subsection{A complete census of accretion activity in nearby galaxies}
The sub-arcsecond resolution and high sensitivity of \emerlin{} enables us to study the non-thermal nuclear emission (jets, disc winds and coronal outflows) \citep[see][]{Panessa2019}. This will enable the study of jet structures and the radio luminosity function of nearby galaxies to sensitivities a magnitude lower than those from previous studies which probe the kinetic feedback of AGN into their host galaxies. This output is intrinsically linked to the accretion process in the X-rays, more sensitive to the BH activity than the radio band, with models predicting a fundamental relationship between the X-ray and radio luminosity with the black hole mass between SMBHs and Galactic X-ray binaries. By resolving out much of the galaxy contaminant radio emission, LeMMINGs will significantly improve the scatter of this part of the `fundamental plane of black hole activity' \citep[e.g.][]{Merloni03,falcke04}.  

\vspace{-0.3cm}
\subsection{The LeMMINGs Sample}

LeMMINGs consists of 280 galaxies sub-selected from the optical Palomar bright spectroscopic catalogue \citep[e.g.,][]{Filippenko85,Ho95,ho97a,ho97b,ho97c,ho97d,Ho03,Ho09}: a statistically-complete sample of 486 galaxies (B$_{T}$ <
12.5 mag) in the northern sky, covering all AGN classes (e.g. Seyfert, LINER, Absorption Line Galaxy, HII galaxy and Transition) and Hubble types (e.g. ellipticals, spirals, lenticulars and irregulars) \citep{BeswickLemmings}. A declination cut of 20$^{\circ}$ was made to the parent sample to prevent elongation of the \emerlin{} synthesized beam. This cut does not reduce the statistical significance of the sample. Being optically-selected, LeMMINGs has no radio bias, and at a median distance of 20 Mpc, reveals the radio natures of our Galactic back yard with sub-kpc linear resolutions.

The survey is split between two samples: shallow and deep. The `shallow' sample comprises all 280 galaxies observed with `snap-shot' observations to optimally distribute \textit{uv}-coverage of up to 48 minutes on source. Each source was grouped with nine others nearby in the sky to make 28 observing blocks for an efficient observing strategy \citep[][]{BeswickLemmings}. The `deep' sample comprises a handful of the 280 galaxies, selected for the breadth of their multi-wavelength data and "interesting" natures to drill-down on specific physics questions pertaining to jet production, SNR rates or galaxy evolution in these sources. Together, these two samples complement one another, with the holistic view of the `shallow' sample informing the `deep' sample, and vice versa.

\vspace{-0.3cm}
\subsection{Observations and Data Reduction}

Observations of the `deep' and `shallow' samples at 1.5 GHz were initially performed in 2015-2017, with 5 GHz `shallow' data following in 2017-2019. Initially observations were manually flagged and calibrated using AIPS \citep{AIPS}, but the development of the \emerlin{} CASA \citep{CASA} Pipeline \citep[eMCP][]{eMCP} has revolutionised the processing of \emerlin{} datasets. All LeMMINGs datasets have now been reduced and re-analysed using the eMCP although only the 1.5 GHz sample has been published so far \citep[see ][for full description of reduction procedures]{BaldiLeMMINGs,BaldiLeMMINGs2,BaldiLeMMINGs3}. The 5\,GHz sample is being analysed and will be presented in future publications alongside the 1.5\,GHz sample.


\section{Initial findings from LeMMINGs}

\vspace{-0.3cm}
\subsection{The `deep' sample}

\begin{figure}
    \centering
    \includegraphics[width=\textwidth]{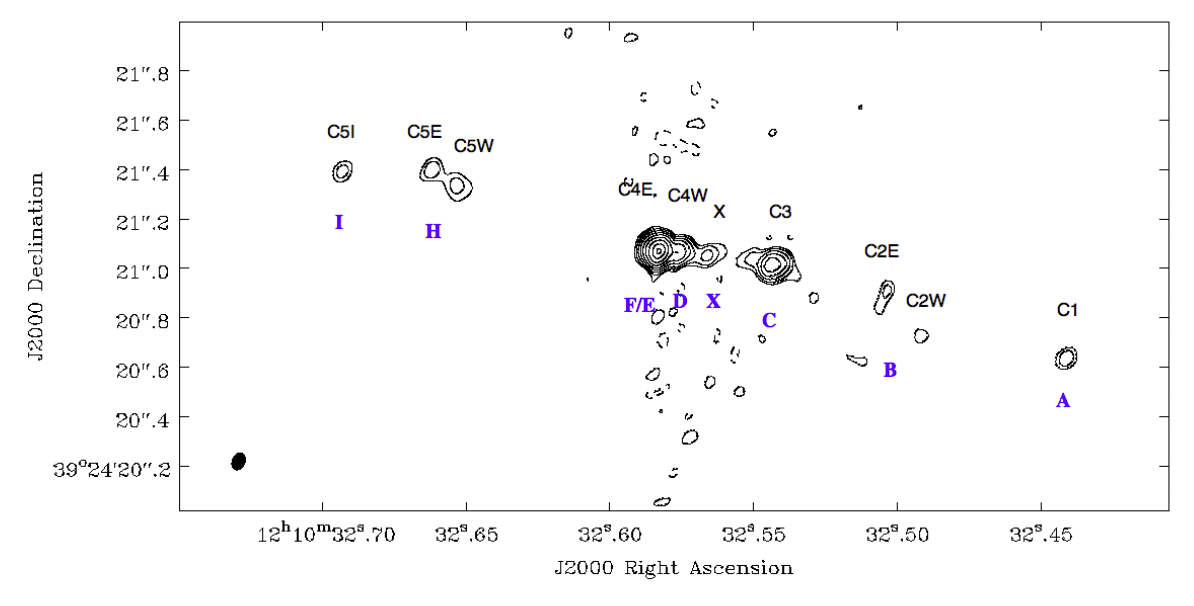}
    \caption{5\,GHz image of NGC 4151 from \citep[][their Fig.~1]{Williams4151_2}, showing the highly resolved radio structure of the inner region of the galaxy. The AGN resides in C4W but the brightest component, C4E is the first radio component in the jet interacting with the extended emission line region in the galaxy.}
    \label{fig:NGC4151}
\end{figure}

The purpose of the `deep' sample was to study individual objects for full-track observations in order to build up sensitivity and \textit{uv}-coverage to focus on specific science projects. In all, six sources have been observed and published as part of these deep observations: M82, a local supernova factory \citep{MuxlowM82}; IC 10, a complex irregular star forming galaxy \citep{Westcott2017}; NGC 4151, a well-known Seyfert galaxy with radio jets \citep{Williams4151,Williams4151_2}; NGC 5322, an elliptical galaxy with collimated large scale radio jets \citep{Dullo5322}; M51, an interacting galaxy with two AGNs \citep{RampadarathM51} and NGC 6217, a star-forming galaxy with a putative X-ray jet \citep{Williams6217}. 

As an example of the usefulness of the `deep' sample, the Seyfert galaxy NGC 4151 had been studied for decades, including with \emerlin{}'s predecessor, MERLIN. Re-visiting NGC 4151 with \emerlin{} full-track imaging enabled deeper observations to pick up low-level radio emission along the jet axis. However, the most interesting aspect of the work showed that the nuclear region had nearly doubled in flux density over a 20 year period \citep{Williams4151}. Follow-up observations at 5 GHz showed that the cause of this increase was the first jet component east of the active nucleus \citep[][see Fig.\ref{fig:NGC4151}]{Williams4151_2}, indicating continued jet interaction with the extended emission line region in NGC 4151. Even with \emerlin{} resolution of 50\,mas at 5 GHz, it is still not possible to unambiguously resolve the AGN from the rest of the jet, showing the difficulty in removing non-AGN-related emission from even well-known and widely-studied sources such as NGC\,4151. 

\vspace{-0.3cm}
\subsection{The `shallow' sample}

The `shallow' sample 1.5\,GHz, at 150\,mas resolution data has been reduced, analysed, imaged and published, with typical r.m.s. noise levels of $\sim$80$\mu$Jy/beam \citep{BaldiLeMMINGs,BaldiLeMMINGs2,BaldiLeMMINGs3}. The 5\,GHz sample will be presented in future papers (see Section~\ref{sec:5GHzsample}). Thus far the 1.5\,GHz shallow sample has focused on the central regions, e.g. the AGN and radio jets, but future works will perform widefield images (see Section~\ref{sec:widefield}) to map out the entire optical extent of all LeMMINGs galaxies searching for off-nuclear and SF related emission. \citep{BaldiLeMMINGs} re-classified the nuclear AGN types for all LeMMINGs galaxies via the updated Baldwin/Phillips/Terlevich (BPT) diagrams \citep{BaldwinPhillipsTerlevich81} from \citep{Kewley06} and \citep{Buttiglione2010} using the optical emission line ratios. Transition galaxies were removed from the classification diagrams in order to give a more robust determination of AGN/non-AGN activity. Dynamical black hole mass measurements from \citep{vanderbosch16} were used if possible. If a reliable mass measurement was not available in the literature, then masses were estimated via the M$-\sigma$ relation \citep[][]{tremaine02} using stellar velocities from the Palomar sample \citep{Filippenko85,Ho95,ho97a,ho97b,ho97c,ho97d,Ho03,Ho09}. Galaxy morphological classifications are included from the literature to compare the connection between host galaxy and nuclear type. 
Below are some of the key science highlights of the LeMMINGs 1.5\,GHz statistical sample to date.

\vspace{-0.3cm}
\subsubsection{Overview of the 1.5\,GHz results}
The 1.5\,GHz programme represents the deepest high-resolution radio survey of nearby galaxies to-date. In total the LeMMINGs 1.5\,GHz radio data detected radio emission $>0.25$\,mJy/beam in 125/280 galaxies in the sample (44.6$\%$) \citep{BaldiLeMMINGs2}, similar to previous radio detection rates of the Palomar sample \citep[e.g.][]{nagar05}. However, these previous radio studies have mostly focused on the LINERs and Seyfert galaxies whereas LeMMINGs includes SF dominated sources, so our detection rate includes many objects that would have been missed from previous studies. Of these 125 detected LeMMINGs galaxies, 106 displayed radio emission coincident within 1.2-arcsec of the optical nucleus, although only a third of these showed clear `jetted' morphologies: a `smoking gun' for an active nucleus. We were able to achieve a 100$\%$ detection rate for AGN down to BH masses of 10$^{7}$M$_{\odot}$, down to luminosities of L$_{\rm radio,core} > 10^{19.8}$~W/Hz, and which showed a transition break of $\sim$10$^{6.5}$M$_{\odot}$ where the radio luminosity becomes more dominated by SF than AGN processes. We also detected radio signatures of SNRs, supernovae, a tidal disruption event in NGC 3690 and some other interesting non-AGN-related radio sources in the central regions of LeMMINGs galaxies.

\vspace{-0.3cm}

\begin{wrapfigure}{r}{0.6\textwidth}
\includegraphics[width=1.0\linewidth, keepaspectratio=true]{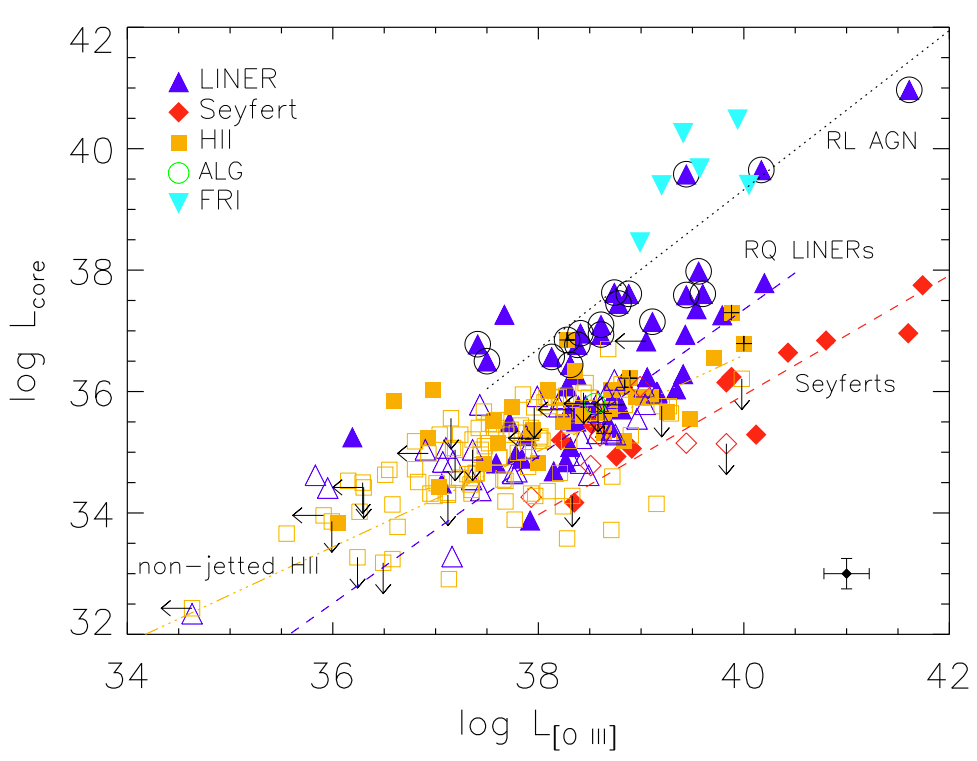}
\caption{Radio vs [\ion{O}{III}] line luminosity diagram from \citep[][their Fig. 6]{BaldiLeMMINGs3}, showing the stratified nature of the correlations between radio-loud AGN from the literature with some LeMMINGs LINERs, radio-quiet LINERs, Seyfect galaxies, and non-jetted star-forming {\text{H\,\textsc{\lowercase{II}}}} galaxies.}
\label{fig:radioOIII}
\end{wrapfigure}

\subsubsection{The `Jetted' {\text{H\,\textsc{\lowercase{II}}}} galaxies}

One of the main goals of LeMMINGs is to discover LLAGN in star forming galaxies by detecting their radio jets. Many star forming galaxies have been missed by previous surveys owing to their lack of clear optical photo-ionisation signatures in BPT diagrams. The 1.5\,GHz sample detected `jetted' radio emission from seven star-forming \ion{H}{II} galaxies, many of which had not been observed before. The `jetted' nature of the radio morphology suggests that these are likely good candidates for an active nucleus. Moreover the multi-wavelength comparisons discussed in the 1.5\,GHz LeMMINGs survey shown that these `jetted' AGN follow similar luminosity correlations to active galaxies e.g. Seyferts and LINERs (e.g. see next Section). Whether or not these are truly AGN or not is yet to be confirmed, but, the 5\,GHz \emerlin{} data (see Section~\ref{sec:5GHzsample}) will provide radio spectral information to help disentangle their nature. Categorising the entire population of nearby LLAGNs is crucial for constructing accurate radio luminosity functions, making the discovery of even a handful of `jetted' \ion{H}{II} galaxies very important.

\vspace{-0.3cm}
\subsubsection{The connection with the [\ion{O}{III}] line luminosity}

The Palomar survey lists [\ion{O}{III}] line luminosities for all of the galaxies, albeit only to 2-arcsec resolution. The [\ion{O}{III}] emission line is known to probe photo-ionisation processes of the nuclear region and therefore is commonly used as a proxy of the AGN bolometric luminosity. LeMMINGs connects the radio emission on the smallest scales with the [\ion{O}{III}] line emission, showing separation in the correlations for different types of source (see Fig~\ref{fig:radioOIII}). Broadly, the most radio-loud LINERs appear to align with radio-loud AGN such as FRI and 3C sources, only at lower luminosities, indicating a continuation of the accretion/outflow paradigm to the lowest levels. Radio-quiet LINERs follow the same gradient but at lower radio luminosities, with Seyferts at a similar gradient but at even lower radio luminosities. The aforementioned 'jetted' \ion{H}{II} galaxies follow the radio-quiet LINERS further indicating that they may be powered by an AGN. This stratification of the accretion/outflow paradigm suggests that broadly the radio and [\ion{O}{III}] line emission are connected down to very low accretion levels, but in different accretion modes.

\vspace{-0.3cm}
\subsection{Multi-wavelength samples}

A key part of the LeMMINGs survey is to compare the \emerlin{} radio data with complementary and wide-reaching multi-wavelength data. Thus far, high resolution X-ray (0.5 arcsec resolution with \textit{Chandra} \citep{XrayLeMMINGs}) data have been published for three quarters of the LeMMINGs sample. It shows a 70$\%$ detection rate of nuclear X-ray sources aligned with the optical galaxy center to a survey limit of 1.65$\times$10$^{-14}$~erg s$^{-1}$ cm$^{-2}$, an increased detection fraction compared to previous works using the same parent sample at lower sensitivity and resolution \citep{RobertsWarwick,HoUlvestad}. While it is possible that some of these sources are X-ray binaries and ultraluminous X-ray sources, further analysis of this data to categorise the timing properties is currently underway (Pahari et al. in prep).

Optical 0.05$-$0.1 arcsec resolution \textit{HST} data (Dullo et al. in prep.) have also been sought. These data can be decomposed into discrete galaxy structures to parameterise the relative contributions of e.g. AGN, bars, and spiral arms. This study is important for linking the AGN or SF activity to galaxy evolution. Further archival datasets with the Infra-red (\textit{Herschel/Spitzer}) will aid in the estimation of SF rates. Further high-resolution VLBI studies of select LeMMINGs samples are underway and low frequency coverage with LOFAR will also be attempted in future.

\section{Future work}
\vspace{-0.3cm}
\subsection{The 5 GHz sample}
\label{sec:5GHzsample}
The 5 GHz sample gives the opportunity to study the same objects at higher frequencies which are less affected by SF processes compared to the 1.5\,GHz sample due to their steeper radio spectral index. The 5\,GHz sample also provides a wider spectral range from which to estimate the spectral index and yields higher resolutions for detecting small scale jets. These three points help with classifying AGN, as they tend to be spectrally-flatter, and compact which leads to higher brightness temperatures being inferred. The deep image of NGC 4151 in Fig~\ref{fig:NGC4151} shows the importance of the 5 GHz data and will enable LLAGN components to be resolved from the large `jetted' structures seen in many of the \emerlin{} 1.5\,GHz images.

\vspace{-0.3cm}
\subsection{Wide-field imaging}
\label{sec:widefield}

The 25m antennas in the \emerlin{} array provides a primary beam at 1.5 and 5 GHz of approximately 30' and 7', respectively. The LeMMINGs galaxies can have angular sizes of several arcminutes so \emerlin{} can provide high-resolution maps of the entire optical extent of all LeMMINGs galaxies to detect off-nuclear emission like SF and {\text{H\,\textsc{\lowercase{II}}}} regions, SNRs and exotic radio sources. 
To do this, developments in widefield imaging software such as \texttt{wsclean} \citep{wsclean,wsclean2} have enabled rapid image deconvolution of the \emerlin{} primary beam at full resolution.

\begin{figure}
    \centering
    \includegraphics[width=\columnwidth]{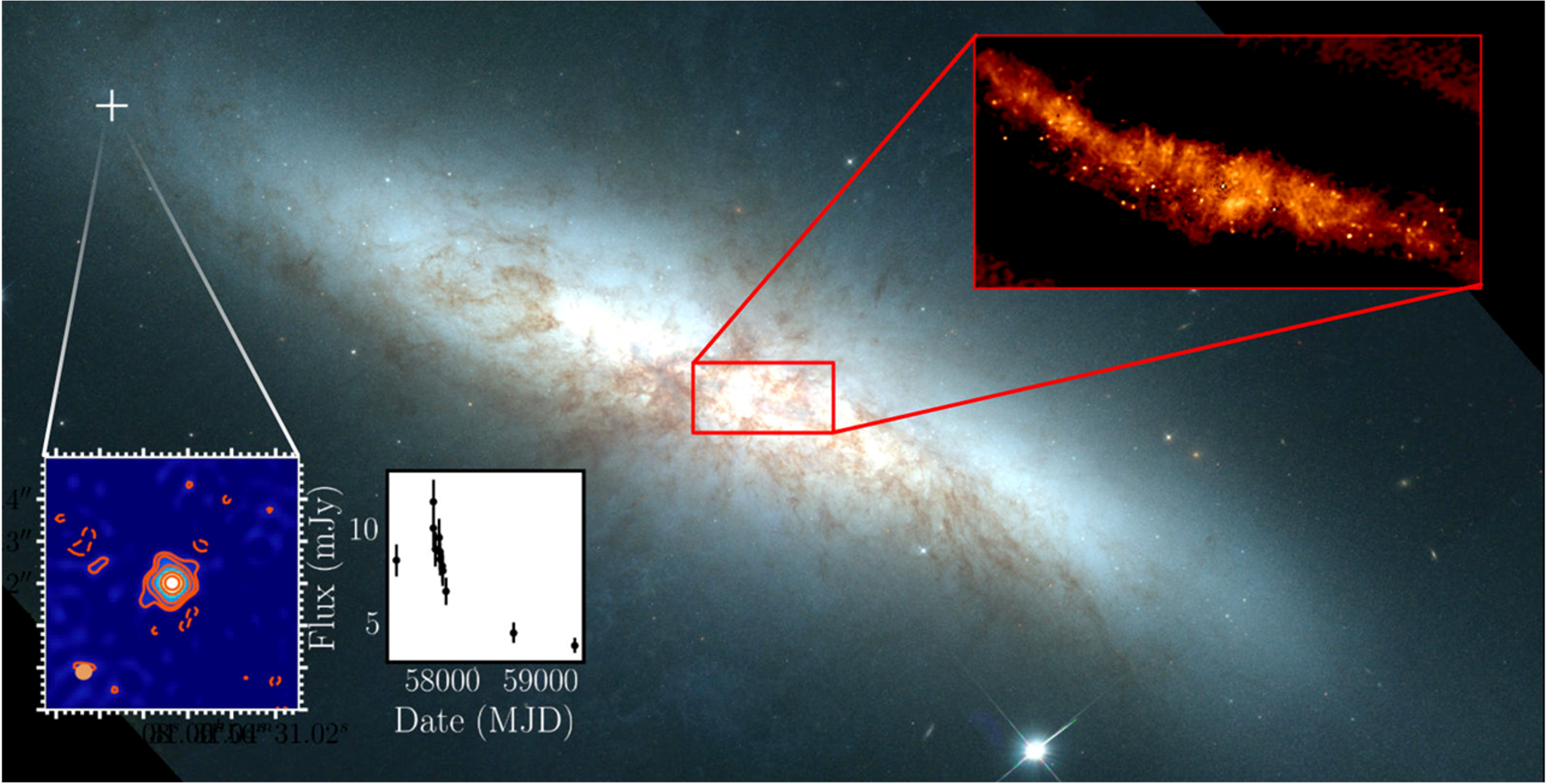}
    \caption{A \textit{HST}  false colour image of M82 with inset \textit{top right}, the radio structure of the central kpc of M82  \citep{MuxlowM82}, and inset \textit{bottom left}, a radio source discovered on the edge of the galaxy with its non-primary-beam corrected light curve from 2015-2019.}
    \label{fig:M82_smudge}
\end{figure}

In Figure~\ref{fig:M82_smudge}, we show an optical image of M82 which includes dozens of SNRs and {\text{H\,\textsc{\lowercase{II}}}} regions in the central kiloparsec \citep[e.g.,][]{MuxlowM82}. However, M82 is $\sim$7-arcmin across, so a large part of M82 has not previously been investigated in the radio at sub-arcsecond resolution. Our preliminary 5 GHz `shallow' data were used to test widefield \textsc{wsclean} algorithms, and detected a new radio source $\sim$4' away from the central pointing, and on the edge of the optical extent of the galaxy.

\vspace{-0.3cm}
\subsection{Further studies of the LeMMINGs sample}
In addition to the wide-field imaging and presentation of the 5\,GHz sample, work is underway to compile radio luminosity functions of the 1.5 and 5\,GHz samples to explore the prevalence of activity at lower luminosities than previous samples. Using the \textit{Chandra} X-ray data with LeMMINGs radio data we will study the radio/X-ray plane of the different types of nuclei present in LeMMINGs. By combining this information with black hole mass estimates, we will compute the fundamental plane of BH activity for a large sample at the highest resolution to-date (Williams et al., in prep.). Eventually, we will also provide all of the radio images and curated data online as a large part of the legacy value of the LeMMINGs survey is the long-term provision of high-resolution multi-wavelength data of nearby galaxies for the wider astronomy community.


\vspace{-0.3cm}
\section{Summary}
\vspace{-0.3cm}
LeMMINGs has only started to scratch the surface of nuclear accretion and star formation processes in nearby galaxies.
Thus far the focus has been on the galaxy central regions, probing the AGN processes at 1.5 GHz; however, 5 GHz data are being analysed, multi-frequency datasets are being compiled, and wide-field imaging techniques are being tested. The dual-frequency approach at high resolution and high sensitivity enables detection of the sub-kpc-jets of the lowest luminosity AGN in our back yard, as well as the mas-scale shells of SNRs.

\vspace{-0.3cm}
\section*{Acknowledgements}
\vspace{-0.3cm}
\emerlin{} is a National Facility operated by the University of Manchester at Jodrell Bank Observatory on behalf of STFC, part of UK Research and Innovation.

\bibliographystyle{JHEP}
\bibliography{skeleton.bib}



\end{document}